\documentstyle[12pt]{article}
\raggedright 
\renewcommand{\section}[1]{\refstepcounter{section}
\vspace{24pt}\noindent{\bf\arabic{section}.\quad #1}
\vspace*{12pt}}
\newcommand{\ulsect}[1]{\vspace{18pt}\noindent{\bf #1}
\vspace*{12pt}}

\begin{document}
\begin{flushright} CERN-TH.6771/93\\
\end{flushright}
\vspace*{5mm}
\begin{center}
{\bf Analysis techniques for high-multiplicity
collisions} \\[10mm]
David Seibert \\[3mm] Theory Division, CERN, CH-1211
Geneva 23, Switzerland$^*$\\[30pt]
{\bf Abstract} \\
\end{center}
I discuss methods for identifying and quantifying
phase transitions in particle collisions, concentrating on two
techniques for use in ultra-relativistic nuclear collisions.
The first technique is to use rapidity correlation measurements
to determine the correlation length, while the second is to use
the transverse mass distribution of dileptons in the
$\rho^0-\omega$ peak to determine the transition temperature.
\vfill
\begin{center}
{\em Talk for the XXIst International Workshop on\\
Gross Properties of Nuclei and Nuclear Excitations\\
Hirschegg, Kleinwalsertal, Austria\\
January 18-23, 1993}
\end{center}
\vspace*{10mm}
\begin{flushleft} CERN-TH.6771/93 \\
January 1993 \\ \end{flushleft}
\thispagestyle{empty}\mbox{}
\footnoterule
{$^*$On leave until October 12, 1993 from: Physics Department,
Kent State University, Kent, OH 44242 USA. Internet address:
seibert@surya11.cern.ch.}
\newpage
\setcounter{page}{1} \pagestyle{plain}

\section{Introduction}

The goal of this talk is to present two techniques that can be
used for quantifying phase transitions in ultra-relativistic
nuclear collisions.  I discuss these techniques in a general
manner, to show how they might be used to study other
collisions and other phase transitions.  I first give a very
general overview of an ultra-relativistic nuclear collision.

The basic physics of an ultra-relativistic nuclear collision is
simple [\ref{rBj}].  First, a lot of partons collide in a proper
time less than about 1 fm/c, producing an enormous number of
secondary particles.  Then, in another 1 fm/c or so, the
secondary particles (hopefully) equilibrate.  I use proper time
rather than time, because time varies with the reference frame,
while the system looks approximately the same in all frames at
equal proper time.

The period after equilibration is potentially the most interesting,
as there is a chance to measure bulk properties of hot hadronic
matter during this period.  Many things might go on here.
Typically, people hope that immediately after equilibration the
hot secondaries will exist as a deconfined quark-gluon plasma
or as chirally-symmetric hadronic matter.  As the matter expands and
cools, there should be some sort of transition, possibly an actual
phase transition, from this exotic state back to normal hadrons.
Finally, as the system expands further the matter will freeze out
and flow to the detectors when the mean free path is long compared
to the size of the system.

For now, I assume that this picture is correct, and that there is a
phase transition to observe.  I assume as little as possible about
the nature of this transition; rather, I discuss ways to observe the
physics of the transition.  I discuss two proposed techniques in
sections \ref{sclm} and \ref{sttm}, and summarize in section \ref{ss}.

\section{Correlation length measurement} \label{sclm}

At a first-order phase transition the correlation length is usually
different for the two coexisting phases, while at a second-order
transition the correlation length diverges at the critical point.
Thus, a change in the correlation length might indicate a phase
transition.  However, correlation lengths are not very easy to
measure in particle collisions for two reasons: i) lengths are
not directly measured but are inferred from velocity (rapidity)
correlation measurements, and ii) the sample size typically
changes from event to event.  For brevity, I discuss only the first
difficulty here.

The correlation length inference is simple in concept.  If the
phase transition is first-order, the transition occurs through a
mixed-phase region, in which regions of the high- and
low-temperature phases coexist.  [In a second-order transition,
there are instead very large correlated regions that form as
the correlation length diverges at the critical temperature.]
As interactions between particles with very different velocities
are weak, the particles produced from these regions will be
correlated in velocity.

The shape of the velocity (or rapidity) correlation function is
fixed in the most likely case that the decays of the regions are
isotropic in their rest frames, giving a scale of 1--2 units of
rapidity (the same as for the correlations seen in pp collisions).
The magnitude of the correlation  function is proportional to the
number of particles emitted per region, so this number can be
obtained directly. The scale of the regions (and hence the mean
volume) is then determined by the correlation length, while the
volume is also related to the number of particles, so the
correlation length can be inferred from the velocity correlations
with the help of some theory [\ref{rrs}].  Useful methods for
measuring the correlations and results from current experiments
are given in Refs.~[\ref{rsbcfs}] and [\ref{rces}].

One background for the fluctuations from a phase transition is
the correlations due to hadronization.  This background is
most easily calculated by extrapolating correlation measurements
from pp reactions, in which hadronization ocurs but no bulk
phase transition is expected.  Current results indicate that the
correlations seen in O+Em and S+Em collisions at $\sqrt{s}=20$
GeV/nucleon are an order of magnitude larger than the signal
extrapolated from pp data [\ref{rcldat}], so this background can
largely be ignored.

It is easier to see large changes in the correlation length
(signalling a phase transition) than to measure the correlation
length, as knowledge of the density is not necessary.  A large
change in the correlation length follows directly from a large
change in the number of particles per region.  However,
looking for a large change is not necessarily easy, and
it gets more difficult as the rapidity density, $dN/dy$,
increases [\ref{rcls}].  The correlation signal is typically
proportional to $(dN/dy)^{-1}$, while the statistical noise is
proportional to $(dN/dy)^{-1/2}$, so the signal-to-noise ratio
decreases as $(dN/dy)^{-1/2}$.  Even worse, there is a
non-statistical background from Bose-Einstein interference that
is approximately independent of $dN/dy$, so the
signal-to-background ratio can decrease as fast as
$(dN/dy)^{-1}$!  Thus, while this technique is simple in
principle, it becomes infinitely difficult (and probably
impossible) in the limit $dN/dy \rightarrow \infty$.

\section{Transition temperature measurement} \label{sttm}

The proposal for measurement of the hadronic transition temperature
is a more elaborate version of a proposal by Siemens and Chin for
detecting the existence of the transition [\ref{rsc}].  The
technique is simple [\ref{rTcs}]: the experimenters measure the
transverse mass, $m_T = (p_T^2+m^2)^{1/2}$, distribution of lepton
pairs in the $\rho^0-\omega$ peak.  This distribution is then fit
to a thermal shape, and the temperature extracted from the fit is
the transition temperature, $T_t$, with theoretical corrections of
order 25\%.  Some of the theoretical corrections are discussed in
Ref.~[\ref{rsmf}].

Dileptons are used (rather than pions, that would allow
separation of $\rho^0$ and $\omega$ mesons) because pions can only
escape from the final state matter, as those formed earlier
interact on the way out and their origin from resonances is
lost.  As leptons are not strongly-interacting, they can escape
from the hot matter at much earlier times, so they are a better
probe of the early stages of the collision.

Of course, once $T_t$ is extracted then there is the more
difficult problem of its interpretation.  Because of the generality
of the method, this will not be very rigorous; essentially, $T_t$
is the temperature at which the $\rho$ mesons disappear.  This
could be due to dynamics, in which case $T_t$ should depend on
collision energy and the sizes of the colliding nuclei, or it could
be due to a bulk phase transition, in which case $T_t$ should be
constant.  Even in the case of a phase transition, there is the
question of the nature of the phase transition.

The technique works for many possible hadronic transitions
(chiral symmetry restoration, deconfinement, etc.) because $\rho$
mesons are {\em order parameters} for all of these transitions.
The $\rho$ exists in the low-temperature phase, but not in the
high-temperature phase.  The signal from $\rho$ mesons is thus an
experimenter's $\theta$-function, measuring properties of the
system below $T_t$.

The $\rho$ mesons are also especially useful because they have a
very short lifetime, $\tau_{\rho} = 1.3$ fm/c.  Their main decay
mode is into two pions, so by detailed balance their short
lifetime guarantees a fast rate for the reaction $\pi \pi
\rightarrow \rho$.  Thus, the $\rho$ mesons remain near
equilibrium (and in good thermal contact with the pions)
whenever the expansion time is more than about one fm/c.

Because the $\rho$ mass is large ($m_{\rho} \gg T_t$), the number
of $\rho$ mesons in the system (and thus the $\rho^0$-peak
dilepton signal) decreases exponentially with $T^{-1}$ as the
system cools below $T_t$.  Most of the signal is produced when the
temperature is near $T_t$, in a band with width of order
$T_t^2/m_{\rho}$ below $T_t$.  Thus, the theoretical corrections
to $T_t$ are of order $T_t^2/m_{\rho}$.  The $rho$ mesons at
freezeout are described fairly well by an equilibrium distribution
at the freezeout temperature, remaining in the system for proper
time $\tau_{\rho}$, so the freezeout signal is small if the system
remains near equilibrium much longer than $\tau_{\rho}$.

Thus, the requirements for use of this technique are:

\begin{list}{\arabic{enumi}.\hfill}{\setlength{\topsep}{0pt}
\setlength{\partopsep}{0pt} \setlength{\itemsep}{0pt}
\setlength{\parsep}{0pt} \setlength{\leftmargin}{\labelwidth}
\setlength{\rightmargin}{0pt} \setlength{\listparindent}{0pt}
\setlength{\itemindent}{0pt} \setlength{\labelsep}{0pt}
\usecounter{enumi}}

\item Some particle must vanish for $T>T_t$ (order parameter).

\item The lifetime of the particle must be short.

\item The mass should be much greater than $T_t$.

\item Observable decay products must escape the collision volume
at early times.

\end{list}

It would appear at first glance that this technique should work
well at all values of $dN/dy$.  The technique is in principle
possible even for $dN/dy \rightarrow \infty$, as the
signal-to-background ratio is constant.  However, the signal is
difficult to measure, as the background (mainly statistical
dileptons from $\pi^0$ decays) is large.

There is one further complication, that there should not be too
many nearby resonances, so that the peak being used to identify
the desired mesons can be extracted from the background.  In
the application to the hadronic phase transitions, this is not
strictly true, as the $\omega$ is nearly degenerate with the
$\rho$.  This complicates the signal, but the background from
$\omega$ mesons is much smaller than the signal from the $\rho^0$
mesons (because of the much faster $\rho$ decay rate) if the
system spends enough time near equilibrium.

\vfill \eject

\section{Summary} \label{ss}

I have presented here the outlines of two techniques for
identifying and characterizing phase transitions in particle
collisions.  The first technique, searching for a large change
in the correlation length, is simple in concept but becomes
impossible even in principle for large $dN/dy$.  This technique
provides a relatively theory-independent method to search for
phase transitions.  The second technique, obtaining the
transition temperature from the $m_T$ spectrum  of dileptons in
the $\rho^0-\omega$ peak, is more complicated in concept but is
possible in principle even in the limit $dN/dy \rightarrow
\infty$.

Both of these techniques should be possible to apply to other
nuclear and high energy systems that might exhibit phase
transitions.  The second technique depends on having a
fortuitous resonance that serves as an order parameter but has
a short lifetime.  This is probably not unlikely for
strongly-interacting systems, due to the large number of
resonances, but may make application to weakly-interacting
systems difficult.  The first technique should be applicable
to the study of almost any transition that occurs at not too
large values of $dN/dy$.

\ulsect{Acknowledgements}

I thank Dr.\ M. Jacob for critical reading of the manuscript,
and the Theory Division of CERN for their hospitality. This
material is based upon work supported by the North Atlantic
Treaty Organization under a Grant awarded in 1991.

\ulsect{References}

\begin{list}{\arabic{enumi}.\hfill}{\setlength{\topsep}{0pt}
\setlength{\partopsep}{0pt} \setlength{\itemsep}{0pt}
\setlength{\parsep}{0pt} \setlength{\leftmargin}{\labelwidth}
\setlength{\rightmargin}{0pt} \setlength{\listparindent}{0pt}
\setlength{\itemindent}{0pt} \setlength{\labelsep}{0pt}
\usecounter{enumi}}

\item J.D. Bjorken, Phys.\ Rev.\ D {\bf 27} (1983) 140.
\label{rBj}

\item P.V. Ruuskanen and D. Seibert, Phys.\ Lett.\ B {\bf 213}
(1988) 227. \label{rrs}

\item S. Voloshin and D. Seibert, Phys.\ Lett.\ B {\bf 249} (1990)
321; D. Seibert and S. Voloshin, Phys.\ Rev.\ D {\bf 443} (1991)
119. \label{rsbcfs}

\item P. Carruthers, H.C. Eggers and I. Sarcevic, Phys.\ Rev.\ C
{\bf 44} (1991) 1629. \label{rces}

\item D. Seibert, Phys.\ Rev.\ C {\bf 44} (1991) 1223.
\label{rcldat}

\item D. Seibert, ``Correlation measurements in high-multiplicity
events,'' Kent State University preprint KSUCNR-011-92 [University
of Minnesota preprint TPI-MINN-92/47-T] (September 1992), submitted
to Phys.\ Rev.\ C. \label{rcls}

\item P.J. Siemens and S.A. Chin, Phys.\ Rev.\ Lett.\ {\bf 55}
(1985) 1266. \label{rsc}

\item D. Seibert, Phys.\ Rev.\ Lett.\ {\bf 68} (1992) 1476.
\label{rTcs}

\item D. Seibert, V.K. Mishra and G. Fai, Phys.\ Rev.\ C {\bf 46}
(1992) 330. \label{rsmf}

\end{list}

\vfill \eject

\end{document}